\documentclass[conference]{IEEEtran}

\IEEEoverridecommandlockouts
\usepackage{cite}
\usepackage{amsmath,amssymb,amsfonts}
\usepackage{algorithm}
\usepackage{algpseudocode}
\usepackage{graphicx}
\usepackage{textcomp}
\usepackage{blindtext}
\usepackage{subfigure}
\usepackage{float} 
\usepackage{xcolor}
\usepackage{diagbox}
\usepackage{slashbox}
\usepackage{wrapfig}
\usepackage{makecell}
\usepackage{graphicx}
\usepackage{booktabs} 
\usepackage[letterpaper, left=0.625in, right=0.625in, top=0.75in, bottom=1.1in]{geometry}
\def\BibTeX{{\rm B\kern-.05em{\sc i\kern-.025em b}\kern-.08em
		T\kern-.1667em\lower.7ex\hbox{E}\kern-.125emX}}
\makeatletter

\begin{document}
	\title{A Disaster-Aware Integrated TN-NTN System-Level Simulator for Resilient 6G Wireless Networks}
	\IEEEpeerreviewmaketitle
	\author{\IEEEauthorblockN{Donglin Wang\IEEEauthorrefmark{2}, Anjie Qiu\IEEEauthorrefmark{2}, Qiuheng Zhou\IEEEauthorrefmark{1}, and Hans D. Schotten\IEEEauthorrefmark{2}\IEEEauthorrefmark{1}}
		\IEEEauthorblockA{\textit{\IEEEauthorrefmark{2}Rhineland-Palatinate Technical University of Kaiserslautern-Landau, Germany} \\
		$\{$dwang,qiu,schotten$\}$@eit.uni-kl.de \\} 
		\IEEEauthorblockA{\textit{\IEEEauthorrefmark{1}German Research Center for Artificial Intelligence (DFKI GmbH), Kaiserslautern, Germany} \\
		$\{$qiuheng.zhou,schotten$\}$@dfki.de}
	}
	\maketitle
\begin{abstract}
Non-terrestrial networks (NTN) have been standardized by the 3rd generation partnership project (3GPP) as a key component of future 6G systems to enhance coverage and resilience. In particular, NTN technologies such as low earth orbit (LEO) satellites, high-altitude platform stations (HAPS), and unmanned aerial vehicles (UAVs) are expected to support terrestrial networks (TN) during extreme events and disasters. In this paper, we present a lightweight system-level simulator for evaluating post-failure fallback behavior in integrated TN–NTN wireless networks under a partial-failure disaster model. The simulator follows 3GPP Rel-17/18 modeling principles, supports probabilistic terrestrial next-generation nodeB (gNB) failures, service migration to NTN. 
The simulator supports comparative analysis of throughput, packet reception ratio (PRR), and latency under different user loads, disaster severities, and NTN provisioning levels. Results show the expected capacity-delay tradeoff of terrestrial operation, the reliability and stability of non-terrestrial service, and the balanced resilience behavior of hybrid TN--NTN operation. The proposed framework provides a tractable tool for studying wireless network resilience and traffic management in future integrated 6G mobile systems.

\end{abstract}


\section{Introduction}
The increasing reliance of modern societies on mobile communication networks makes their resilience to natural disasters and large-scale failures a critical design requirement. Terrestrial cellular networks (TN) are particularly vulnerable to infrastructure damage, power outages, and congestion during emergencies. To address these challenges, 3GPP has introduced non-terrestrial networks (NTN) as an integral part of 5G-Advanced and future 6G systems~\cite{3gpp38811,3gpp38821}. NTN technologies, including low-earth orbit (LEO) satellites, high-altitude platform stations (HAPS), and unmanned aerial vehicles (UAV)-based gNBs, offer wide-area coverage and rapid deployment capabilities. However, evaluating the joint behavior of TN and NTN during disaster scenarios requires dedicated simulation tools that capture network coupling, service migration, and recovery dynamics. Existing studies often focus either on high-fidelity physical-layer modeling or on isolated TN or NTN performance~\cite{kodheli2021,giordani2020}.

System-level simulation of cellular networks has been widely studied in the literature, with standardized channel and deployment models defined in 3GPP TR~38.901~\cite{3gpp38901}. NTN extensions have been introduced in TR~38.811 and TR~38.821, focusing on propagation effects, latency, and coverage characteristics\cite{3gpp38811,3gpp38821}.
Recent survey literature highlights NTN as a key enabler for resilient 6G coverage and emergency communications, while identifying open challenges in multi-layer resource orchestration, mobility management, and scalable system evaluation \cite{kodheli2021, giordani2020}. 

Recent literature highlights a paradigm shift toward 3D wireless networks to enhance disaster resilience. Alhammadi et al. \cite{alhammadi2024envisioning} demonstrate the feasibility of integrating terrestrial, aerial, and satellite segments to improve situational awareness, while identifying critical challenges in cybersecurity and regulatory coordination. Expanding on autonomous recovery, Yagan et al. \cite{yagan2024fast} introduce Open6GRAN, a 6G framework utilizing reconfigurable intelligent surfaces (RIS), cell-free massive MIMO (cfMIMO), and joint communication and sensing (JCAS) under artificial intelligence (AI) orchestration to ensure uninterrupted service through rapid reconfiguration. To optimize these multi-layered NTNs, Koç et al. \cite{kocc2025score} propose a score-based user allocation (SUA) method that dynamically assigns network elements based on latency, data rate, and mobility demands.

While existing research focuses on complex 6G frameworks and distributed satellite architectures that require significant computational overhead and regulatory coordination, these works often struggle to satisfy high user demands in chaotic post-disaster environments. In contrast, our simulator provides a new lightweight, parameter-driven system-level engine for rapid "what-if" experimentation, focusing on the longitudinal phases of a disaster from recovery failure. By explicitly modeling the coupling between terrestrial and satellite links, my work identifies critical core bottlenecks and congestion factors that high-fidelity specialized tools may overlook.

\begin{figure}[htp]
\centering
\includegraphics[width=\columnwidth]{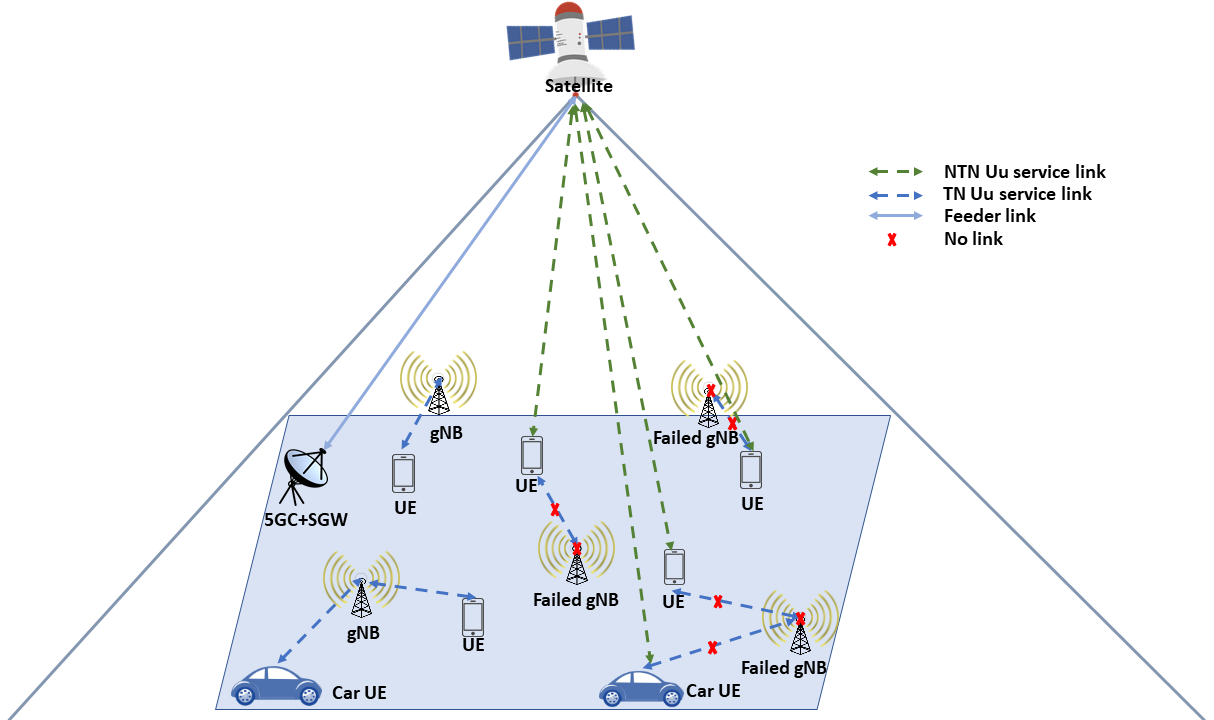}
\caption{Overall integrated TN--NTN simulation architecture for post-disaster evaluation.}
\label{fig:scenario_architecture}
\end{figure}

\section{System Architecture and Scenario Description}

The heterogeneous network architecture under investigation is illustrated in Fig.~\ref{fig:scenario_architecture}. The framework integrates a conventional TN and an NTN overlay to study service continuity under terrestrial failure and controlled offloading.

\subsection{Architectural Components}
\begin{itemize}
    \item \textbf{Space Segment:} The LEO satellite layer provides wide-area overlay coverage and serves as an alternative access layer when terrestrial nodes are degraded or unavailable.
    \item \textbf{Terrestrial Segment:} Multiple 5G/6G next-generation nodeBs (gNBs) are deployed to facilitate high-capacity, low-latency communication for local UE.
    \item \textbf{Ground Infrastructure:} An NTN gateway (GW) manages the \textit{Feeder Link}, providing the necessary backhaul interface between the satellite and the core network infrastructure.
\end{itemize}

\subsection{Interfaces and Link Dynamics}
The connectivity within this heterogeneous environment is governed by several critical interfaces:

\begin{itemize}
\item \textbf{Uu Service Link:} The radio interface between the UEs and the network nodes (both gNBs and the satellite) is defined as the 3GPP-standardized \textit{Uu interface}. This interface enables dynamic selection between terrestrial and non-terrestrial access links based on real-time context.
\item \textbf{Heterogeneous Access:} The architecture supports diverse terminals, including mobile handsets and vehicular units (Car UE), and evaluates their access to either terrestrial or non-terrestrial service depending on the operating condition.
\item \textbf{Resilience and Service Migration:} Terrestrial link failures are represented by the red markers ($\mathbf{\times}$), simulating a disaster scenario where some terrestrial gNBs are offline. In this context, affected traffic is migrated from the degraded terrestrial link to the satellite service link according to the configured TN/NTN split and load conditions, thereby preserving service continuity at system level.
\end{itemize}

\subsection{Simulation Area and Deployment Geometry}
The simulation considers an urban disaster scenario corresponding to a medium-sized city district in Germany \cite{wagner2019regional}. The TN is deployed over a square area, which is consistent with system-level evaluation practices in 3GPP TR~38.901 for urban macro-environmental scenarios (UMa) \cite{3gpp38901}.

\subsection{Terrestrial Network Model}
The TN model comprises $N_{\mathrm{gNB}}=10$ gNBs in a perturbed hexagonal layout (inter-site distance (ISD) $\approx 500$ m) to simulate realistic inter-cell interference. UEs are uniformly distributed and associated with gNBs via a load-aware selection rule. This mechanism penalizes heavily loaded sites, capturing the impact of cell congestion and resource availability on user association.

Large-scale fading follows the Okumura--Hata model \cite{hata1980empirical} with 8 dB shadowing. The instantaneous SINR is calculated by combining received power with thermal noise and aggregate system-level interference, avoiding heavy waveform simulations while preserving inter-cell coupling. Key parameters are detailed in Table \ref{tab:tn_simulation_parameters}.

\begin{table}[h]
\centering
\caption{Terrestrial network simulation parameters.}
\label{tab:tn_simulation_parameters}
\begin{tabular}{|l|c|}
\hline
Parameter & Value \\
\hline
Carrier frequency & 3.5 GHz \\
\hline
Bandwidth & 20 MHz \\
\hline
Number of gNBs & 10 \\
\hline
Layout & Hexagonal with perturbation \\
\hline
Inter-site distance (approx.) & 500 m \\
\hline 
Path loss model & Okumura--Hata \\
\hline
gNB Tx power & 43 dBm \\
\hline
Shadow fading std. dev. & 8 dB \\
\hline
Noise figure & 3 dB \\
\hline
Traffic model & Poisson arrivals \\
\hline
\end{tabular}
\end{table}

Per-user throughput is derived from an implementation-efficient Shannon capacity based on shared bandwidth. End-to-end (E2E) latency is modeled as the aggregate of propagation, processing, and load-dependent queuing delays. This captures the primary effects of disaster-related congestion: reduced throughput due to bandwidth sharing and increased latency from queue buildup at surviving sites.

\subsection{Non-Terrestrial Network Model}
The NTN is modeled as a LEO satellite overlay at 600 km altitude, operating at 2 GHz to provide wide-area coverage independent of the TN. Following 3GPP Rel-17 \cite{3gpp38811, 3gpp38821}, propagation is governed by free-space path loss (FSPL) and slant-range-dependent delay. With a 10$^\circ$ minimum elevation angle, the footprint ($>4000$ km diameter) ensures total service continuity over the terrestrial simulation area. UEs associate with the satellite providing maximum received power, with others contributing to aggregate interference. All the NTN simulation parameters are shown in Table \ref{tab:ntn simulation parameters}.

\begin{table}[h]
\centering
\caption{Non-terrestrial network simulation parameters.}
\label{tab:ntn simulation parameters}
\begin{tabular}{|l|c|}
\hline
Parameter & Value \\
\hline
Coverage & $> 4000$ km diameter\\
\hline
Carrier frequency & 2.0 GHz \\
\hline
Bandwidth & 20 MHz \\
\hline
Orbit type & LEO Satellite \\
\hline
Altitude & 600 km \\ 
\hline
Number of satellites & 10 \\
\hline
Minimum elevation angle & 10$^\circ$ \\
\hline
Path loss model & Free-space \\
\hline
Propagation speed & Speed of light \\
\hline
\end{tabular}
\end{table}

\subsubsection{Architecture and Topology}
We adopt a deterministic non-geostationary (NGSO) topology utilizing a transparent (bent-pipe) payload architecture \cite{3gpp.38.300.Rel17}. Unlike regenerative systems—which process signals onboard for lower latency and higher resilience—the transparent mode relays signals to ground gateways (GWs) for higher-layer processing. This assumption aligns with current 3GPP standards, where disaster resilience is predicated on the survival of the core network and NTN GWs, even if terrestrial radio access nodes fail.

\subsection{Integrated TN--NTN Operation}
The integrated TN--NTN framework is modeled as a dual-layer access system in which TN provides the primary low-latency service and NTN acts as an overlay for resilience support. The two layers are assumed to operate on separate carrier resources; hence, inter-layer radio-frequency interference is neglected. Let the total user set be $\mathcal{U}$, with $|\mathcal{U}| = K$. The users are partitioned into terrestrial and non-terrestrial subsets,
\begin{equation}
\mathcal{U} = \mathcal{U}_{\mathrm{TN}} \cup \mathcal{U}_{\mathrm{NTN}}, 
\qquad
\mathcal{U}_{\mathrm{TN}} \cap \mathcal{U}_{\mathrm{NTN}} = \emptyset,
\end{equation}
where the realized NTN fallback ratio is
\begin{equation}
\hat{\eta} = \frac{|\mathcal{U}_{\mathrm{NTN}}|}{K}.
\end{equation}
In this paper, the configured offload ratio $\eta \in [0,1]$ is an input control parameter that determines the aggressiveness of proactive TN$\rightarrow$NTN migration, whereas the fallback ratio $\hat{\eta}$ is the realized output fraction of users actually served by NTN after gNB failure, forced handover, and post-failure reassociation. Thus, $\eta$ is fixed by configuration, while $\hat{\eta}$ is dynamic and run-dependent. In disaster mode, the partition is not assigned randomly. Instead, a nominal pre-failure TN attachment is computed first, after which users attached to failed gNBs are forcibly moved to NTN. The remaining users are then evaluated against the same degraded TN snapshot, and the provisional terrestrial score of user $u$ is defined as
\begin{equation}
\Gamma_u^{\star} = \max_{b \in \mathcal{B}_{\mathrm{act}}} \Gamma_{u,b},
\end{equation}
where $\Gamma_{u,b}$ is the load-aware association score and $\mathcal{B}_{\mathrm{act}}$ is the active gNB set after failure sampling. Additional proactive NTN offloading is then applied only to weak or overloaded surviving TN users, with the configured aggressiveness $\eta$ controlling how many of these vulnerable users are switched. This two-stage partition ensures that NTN fallback is driven first by actual disconnection and only second by proactive congestion relief on the post-failure terrestrial snapshot.

For each user $u$, the simulator computes rate $R_u$, latency $L_u$, and reception success indicator $\mathbb{I}_u$ based on the serving layer, where
\begin{equation}
\mathbb{I}_u =
\begin{cases}
1, & \mathrm{SINR}_u > \tau,\\
0, & \text{otherwise},
\end{cases}
\end{equation}
and $\tau$ denotes the reception threshold. The aggregate performance is then given by
\begin{equation}
R_{\mathrm{sys}} = \sum_{u \in \mathcal{U}_{\mathrm{TN}}} R_u + \sum_{u \in \mathcal{U}_{\mathrm{NTN}}} R_u,
\end{equation}
\begin{equation}
\mathrm{PRR}_{\mathrm{sys}} = \frac{1}{K}\sum_{u \in \mathcal{U}} \mathbb{I}_u,
\qquad
L_{\mathrm{sys}} = \frac{1}{K}\sum_{u \in \mathcal{U}} L_u.
\end{equation}
Under disaster conditions, terrestrial failures reduce the number of active gNBs and increase TN congestion, which naturally increases the realized fallback ratio $\hat{\eta}$. Small values of the configured aggressiveness $\eta$ preserve TN capacity and low delay by limiting proactive NTN transfers, whereas large values of $\eta$ move more vulnerable users to NTN once failures occur. In the current implementation, the fallback policy is evaluated on a static degraded snapshot and uses a simplified overhead model: NTN-served users incur a fixed migration penalty and the NTN side is subject to a shared feeder-capacity cap with additional latency near saturation, while TN users that reattach to a different surviving gNB incur a lightweight reassociation penalty. In this way, the integrated model captures the essential tradeoff between terrestrial efficiency and non-terrestrial resilience while remaining computationally tractable.


\subsection{Disaster Scenario Model}
The disaster evaluation is modeled as a static post-failure snapshot rather than a full time-evolving recovery process. All the parameters are found in Table \ref{tab:disaster-snapshot parameters}. For each Monte Carlo run, the simulator performs the following steps:
\begin{enumerate}
\item \textbf{Shared Geometry Initialization:} A common UE population is generated once using reference point group mobility (RPGM) panic-aware mobility  \cite{musolesi2006mobility}, and one TN gNB deployment is generated for the same run.
\item \textbf{Failure Sampling:} Each terrestrial gNB is independently marked active or failed according to the failure probability $p_f$, resulting in the post-disaster active set $\mathcal{B}_{\mathrm{act}}$.
\item \textbf{Pre-Failure Attachment and Failure Sampling:} A nominal TN attachment is first computed, and each terrestrial gNB is independently marked active or failed according to the failure probability $p_f$, resulting in the post-disaster active set $\mathcal{B}_{\mathrm{act}}$.
\item \textbf{Forced NTN Handover:} Users attached to failed serving gNBs are treated as disconnected and are forced onto NTN fallback.
\item \textbf{Post-Failure TN Re-Evaluation:} Remaining users are rescored against the degraded TN snapshot using the load-aware association rule.
\item \textbf{Proactive Offloading:} Weak or overloaded surviving TN users are additionally offloaded to NTN, with the configured aggressiveness $\eta$ controlling how many such users are proactively switched.
\item \textbf{Fallback Overhead and KPI Aggregation:} TN metrics are recomputed for the retained TN users, NTN metrics are computed for the fallback users, and simplified TN reassociation penalties together with NTN migration and feeder-capacity constraints are applied before aggregating system-level throughput, packet reception ratio (PRR), and latency across both subsets.
\end{enumerate}

This formulation captures the immediate service-level impact of a partial terrestrial failure and the mitigating role of NTN fallback, while deliberately omitting explicit recovery dynamics inside one simulation run.

\begin{table}[h]
\centering
\caption{Static disaster-snapshot parameters.}
\label{tab:disaster-snapshot parameters}
\begin{tabular}{|l|c|}
\hline
Parameter & Value \\
\hline
Disaster mobility model & RPGM panic \\
\hline
gNB failure probability & 0.5  \\
\hline
Configured offload ratio $\eta$ & 0.5 (proactive aggressiveness) \\
\hline
Realized fallback ratio $\hat{\eta}$ & Dynamic output, $\hat{\eta}=|\mathcal{U}_{\mathrm{NTN}}|/K$ \\
\hline
Offload rule & \makecell[l]{Failed UEs forced to NTN; \\ additional overloaded TN users \\ switched according to $\eta$ }\\
\hline
TN reassociation throughput penalty & 2\% \\
\hline
TN reassociation latency penalty & 1 ms \\
\hline
Migration throughput penalty & 15\% \\
\hline
Migration latency penalty & 3 ms \\
\hline
NTN feeder capacity & 450 Mbps \\
\hline
NTN feeder latency weight & 4 ms \\
\hline
\end{tabular}
\end{table}

\subsection{Disaster-triggered TN--NTN Operation Algorithm}
Algorithm~\ref{alg:tn_ntn_disaster} summarizes the static post-disaster workflow of the proposed TN--NTN simulator. A common UE population and TN deployment are first generated for the run, after which a nominal pre-failure TN attachment is established and a probabilistic terrestrial failure event reduces the active gNB set. During terrestrial association, each user evaluates all candidate gNBs through a load-aware metric rather than pure SINR maximization. Specifically, for user $u$ and candidate gNB $b$, the simulator computes
\begin{equation}
\Pi_b = 20 \log_{10}\!\left(1 + \frac{L_b}{L_{\max}}\right),
\end{equation}
where $L_b$ denotes the current load of gNB $b$ and $L_{\max}$ is the nominal maximum supported load per gNB. The association score is then defined as
\begin{equation}
\Gamma_{u,b} = \mathrm{SINR}_{u,b} - \Pi_b,
\end{equation}
and user $u$ is attached to
\begin{equation}
b^{\star}(u) = \arg\max_{b \in \mathcal{B}_{\mathrm{act}}} \Gamma_{u,b}.
\end{equation}
This mechanism makes overloaded terrestrial cells less attractive and therefore captures congestion-aware access selection at system level. Users disconnected by failed serving gNBs are first forced onto NTN fallback, after which additional weak or overloaded TN users may be proactively switched according to the configured aggressiveness $\eta$. TN metrics are recomputed on the retained TN subset, NTN metrics are evaluated for the fallback subset on the same run, and simplified reassociation, migration, and feeder penalties are applied before system-level aggregation. The algorithm is intentionally abstract and does not model detailed signaling or protocol-accurate handover timing; instead, it captures the dominant resilience mechanisms required for comparative system-level evaluation of a post-failure fallback snapshot.

\begin{algorithm}[!t]
\caption{Static Disaster TN$\rightarrow$NTN Dynamic Handover Workflow}
\label{alg:tn_ntn_disaster}
\begin{algorithmic}[1]
\Require TN parameters $\Theta_{\text{TN}}$, NTN parameters $\Theta_{\text{NTN}}$, number of users $K$, failure probability $p_f$, configured proactive offload aggressiveness $\eta \in [0,1]$
\Ensure KPI metrics (throughput, PRR, latency) for TN, NTN, and the combined degraded snapshot
\State Generate one common UE population $\mathcal{U}$ using panic mobility
\State Generate one TN gNB layout $\mathcal{B}$
\State Compute nominal pre-failure TN attachment for all users
\State Sample the active gNB set $\mathcal{B}_{\mathrm{act}} \subseteq \mathcal{B}$ with failure probability $p_f$
\State Initialize NTN geometry for the same UE population
\State Force users attached to failed serving gNBs onto NTN fallback
\State Compute $\mathrm{SINR}_{u,b}$ for remaining candidate TN users and active $b \in \mathcal{B}_{\mathrm{act}}$
\State Compute load penalty $\Pi_b = 20 \log_{10}\!\left(1 + \frac{L_b}{L_{\max}}\right)$ for each active gNB
\State Form association score $\Gamma_{u,b} = \mathrm{SINR}_{u,b} - \Pi_b$
\State Compute provisional best score $\Gamma_u^{\star} = \max_{b \in \mathcal{B}_{\mathrm{act}}} \Gamma_{u,b}$ for each user
\State Proactively offload weak or overloaded users according to the configured aggressiveness $\eta$
\State Assign the remaining users to $\mathcal{U}_{\mathrm{TN}} = \mathcal{U} \setminus \mathcal{U}_{\mathrm{NTN}}$
\State Recompute TN SINR, rate, and latency for $\mathcal{U}_{\mathrm{TN}}$ on the same degraded snapshot
\State Compute NTN SINR, rate, and latency for $\mathcal{U}_{\mathrm{NTN}}$
\State Apply simplified TN reassociation penalties, NTN migration penalties, and feeder-capacity constraint
\State Combine KPIs:
\Statex \quad $R_{\mathrm{sys}} = \sum_{u \in \mathcal{U}_{\mathrm{TN}}} R_u + \sum_{u \in \mathcal{U}_{\mathrm{NTN}}} R_u$
\Statex \quad $\mathrm{PRR}_{\mathrm{sys}} = \frac{1}{K}\sum_{u \in \mathcal{U}} \mathbb{I}(\mathrm{SINR}_u > \tau)$
\Statex \quad $L_{\mathrm{sys}} = \frac{1}{K}\sum_{u \in \mathcal{U}} L_u$
\State \Return snapshot-level TN/NTN/system KPIs
\end{algorithmic}
\end{algorithm}

\section{Implementation and Metrics}
The simulator is implemented as a parameter-driven Python framework that supports repeatable command-line execution. Monte Carlo simulations are used to capture stochastic geometry, terrestrial failure sampling, and random shadowing effects. The paper reports packet reception ratio, average throughput, and average latency as the primary performance metrics, while auxiliary indicators such as SINR percentiles, outage rate, and gNB load are retained for internal diagnosis as shown in Table \ref{tab:kpis}.

\begin{table}[ht]
\centering
\caption{Key performance metrics generated by the simulator.}
\label{tab:kpis}
\begin{tabular}{|l|l|}
\hline
Metric & Description \\
\hline
SINR & Per-user signal-to-interference-plus-noise ratio \\
\hline 
Average throughput & Mean user data rate \\
\hline
Average latency & Mean end-to-end latency\\
\hline
PRR & \makecell[l]{Fraction of users whose SINR \\ exceeds the reception threshold}\\
\hline
Fallback ratio $\hat{\eta}$ & \makecell[l]{Realized post-failure fraction of users \\ served by NTN fallback}\\
\hline
\end{tabular}
\end{table}



\section{Results and Discussion}
\label{sec:results}

\subsection{Model Validation and Sanity Checks}
The proposed framework is a system-level exploratory simulator rather than a calibrated protocol-stack implementation, so validation is limited to internal sanity checks. The TN model exhibits the expected load-driven throughput reduction and latency growth~\cite{shannon1948,little1961,kleinrock1975}, while the NTN model shows higher per-user throughput with larger constellations and nearly unchanged propagation-dominated latency. In disaster mode, the simulator produces zero realized NTN fallback when $p_f=0$ and converges to the NTN-limited extreme when all gNBs fail, with the fallback ratio increasing monotonically with $p_f$. These checks support internal consistency, although they do not replace external validation against measurements or a reference simulator.

\begin{figure}[!t]
\centering
\includegraphics[width=\columnwidth]{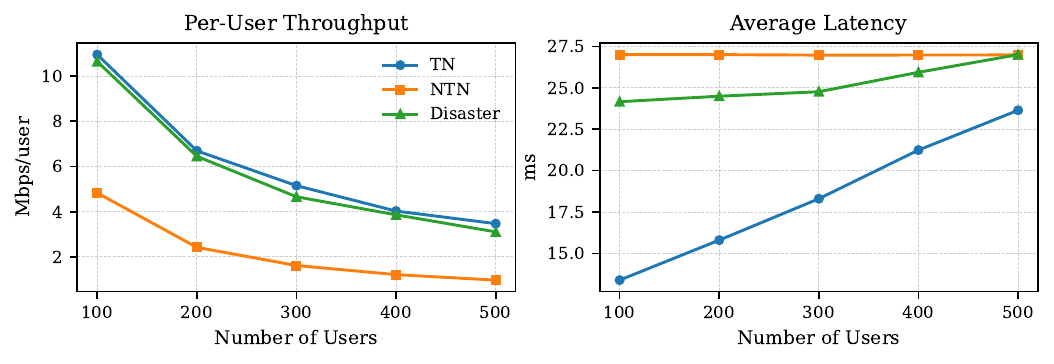}
\caption{Comparison of nominal TN, nominal NTN, and the corrected post-failure dynamic TN--NTN fallback mode in terms of per-user throughput and average latency, including simplified reassociation, migration, and NTN feeder-capacity penalties.}
\label{fig:mode_comparison}
\end{figure}

\subsection{Quantitative Trends in TN, NTN, and Disaster Mode}
Fig.~\ref{fig:mode_comparison} compares nominal TN-only operation, nominal NTN-only operation, and the dynamic post-failure hybrid TN--NTN fallback mode. The hybrid curve corresponds to a disaster configuration with configured proactive aggressiveness $\eta=0.5$ and gNB failure probability $p_f=0.5$, for which the realized NTN fallback ratio varies across runs and user loads.

The simulator was evaluated for user loads from 100 to 500 UEs with 50 Monte Carlo runs per operating point. For nominal TN operation, per-user throughput decreases from 10.952 to 3.471~Mbps/user, while average latency increases from 13.385 to 23.640~ms, which is consistent with finite shared radio resources and queueing under higher load~\cite{shannon1948}. For nominal NTN operation, per-user throughput decreases from 4.833 to 0.971~Mbps/user, while latency remains nearly constant at about 27~ms because the current abstraction is dominated by propagation and fixed processing delay.

The corrected disaster mode exhibits the intended hybrid tradeoff. As load increases, per-user throughput decreases from 10.645 to 3.106~Mbps/user, latency rises from 24.158 to 27.000~ms, and PRR stays in the range \(0.850{-}0.880\). The average number of active terrestrial gNBs remains about 4.6--5.3, while the realized NTN fallback ratio increases from 0.51 to 0.60. Compared with nominal TN, the disaster-mode throughput is consistently lower once explicit failure-driven handover, reassociation overhead, migration penalty, and feeder-capacity limitation are included. Overall, TN remains the most latency-efficient mode under nominal conditions, NTN provides continuity with stable delay but lower capacity, and the hybrid fallback mode offers an intermediate resilience tradeoff.

\subsection{Disaster Severity Sweep Results}
\begin{figure}[!t]
\centering
\includegraphics[width=\columnwidth]{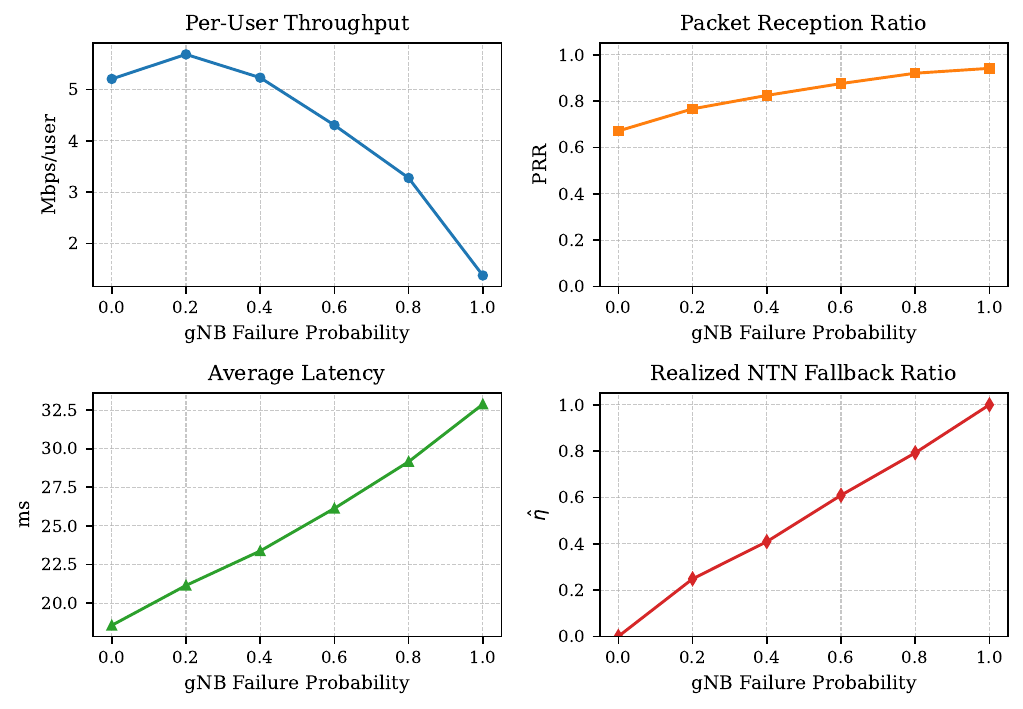}
\caption{Impact of gNB failure probability on post-disaster TN--NTN fallback performance at 300 UEs. Increasing failure severity raises the realized NTN fallback ratio, reduces per-user throughput, increases latency, and improves PRR.}
\label{fig:failure_prob_sweep}
\end{figure}

Fig.~\ref{fig:failure_prob_sweep} shows the effect of terrestrial gNB failure probability at 300 UEs. As \(p_f\) increases from 0 to 1, the realized NTN fallback ratio grows from 0 to 1, per-user throughput first rises slightly from 5.208 to 5.687~Mbps/user at \(p_f=0.2\) because moderate fallback relieves TN congestion, and then drops to 1.375~Mbps/user at \(p_f=1\) when the system becomes fully dependent on the migration- and feeder-limited NTN path. Over the same sweep, PRR increases from 0.672 to 0.942 while latency rises from 18.57 to 32.87~ms, exposing the central tradeoff of the disaster model: higher failure severity increases service continuity through NTN fallback, but at the cost of capacity and delay.

\subsection{NTN Feeder-Capacity Sweep Results}
\begin{figure}[!t]
\centering
\includegraphics[width=\columnwidth]{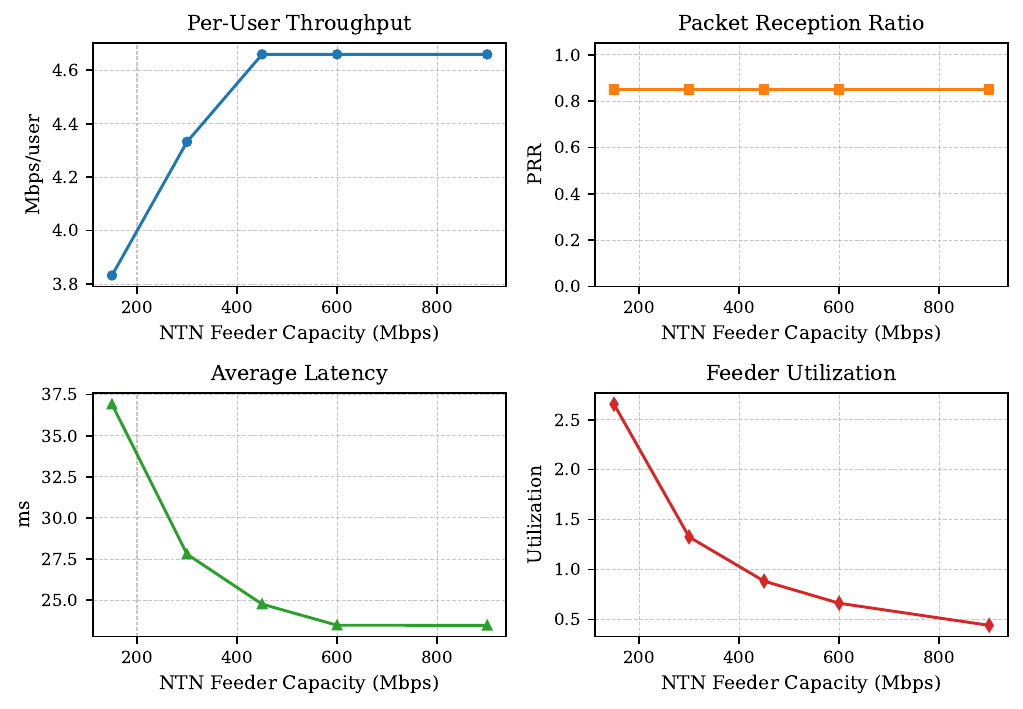}
\caption{Impact of NTN feeder capacity on post-disaster TN--NTN fallback performance at 300 UEs. Throughput and latency improve markedly up to about 450~Mbps, after which the fallback path is no longer feeder-limited.}
\label{fig:feeder_capacity_sweep}
\end{figure}

Fig.~\ref{fig:feeder_capacity_sweep} studies NTN feeder capacity at a fixed disaster operating point (\(p_f=0.5\), \(\eta=0.5\), 300 UEs). Because the realized fallback ratio stays near \(\hat{\eta}\approx 0.513\), this sweep isolates the feeder bottleneck: increasing capacity from 150 to 450~Mbps improves per-user throughput from 3.832 to 4.659~Mbps/user and reduces average latency from 36.91 to 24.76~ms, while PRR remains near 0.850. The weak PRR sensitivity is expected in the current abstraction because PRR is still defined by the access-link SINR threshold, whereas the feeder constraint is modeled as a post-access throughput cap with an added latency term; it therefore degrades delivered rate and delay, but does not materially change whether a packet is counted as successfully received on the radio link. Beyond 450~Mbps, throughput saturates and latency changes little, showing that fallback performance is then limited by access-side sharing and migration overhead rather than by additional feeder scaling.

\section{Conclusion}
This paper presented a disaster-aware, system-level simulator for integrated TN--NTN operation under a partial-failure wireless-network scenario. The framework combines probabilistic terrestrial failures, dynamic TN-NTN handover, simplified TN reassociation and NTN migration overhead, and feeder-limited fallback within a reproducible Monte Carlo evaluation pipeline. Results under increasing user load show the expected capacity--delay tradeoffs in TN, the reliability and delay stability of NTN within the adopted abstraction, and a corrected hybrid disaster mode that no longer dominates nominal TN throughput once explicit handover-related penalties are included, while still preserving a clear PRR advantage under failure. Sensitivity analyzes further indicate that constellation density primarily affects capacity, that disaster severity directly drives the realized NTN fallback ratio, and that feeder provisioning is a dominant bottleneck only below a moderate capacity threshold.
Overall, the simulator provides a transparent platform for exploratory what-if analysis before moving to higher-fidelity tools or external validation benchmarks, and under the stated assumptions TN remains advantageous for low-delay service under moderate load, NTN contributes continuity under terrestrial access degradation, and hybrid operation offers a practical capacity--reliability--delay tradeoff. Future work will extend the present static-snapshot abstraction toward time-varying satellite dynamics, richer failure-and-recovery processes, and broader NTN architectures.

\section{Acknowledgement}
This work has been supported by the Federal Ministry of Education and Research of the Federal Republic of Germany (BMBF) as part of the Open6GHub+ project with funding number 16KIS2406. The authors would like to express their appreciation for the contributions of all Open6GHub+ partners. The authors alone are responsible for the content of the paper, which does not necessarily represent the project.

\bibliographystyle{IEEEtran}
\bibliography{references}

\end{document}